\documentclass{nature}
\usepackage{graphicx}
\usepackage{dcolumn}
\usepackage{bm}
\usepackage{amssymb}
\usepackage{amsmath}
\usepackage{color}
\usepackage{framed}
\usepackage{enumerate}
\usepackage{multirow}

\newcommand{\ket}[1]{\mbox{$\left| #1 \right\rangle$}}

\title{Quantum random number generation}

\author{Xiongfeng Ma$^1$, Xiao Yuan$^{1}$, Zhu Cao$^1$, Bing Qi$^{2, 3}$ and Zhen Zhang$^1$}

\begin{document}

\maketitle
\begin{affiliations}
\item
Center for Quantum Information, Institute for Interdisciplinary Information Sciences, Tsinghua University, Beijing 100084, China.

\item
Quantum Information Science Group, Computational Sciences and Engineering Division, Oak Ridge National Laboratory, Oak Ridge, TN 37831-6418, USA

\item
Department of Physics and Astronomy, University of Tennessee, Knoxville, Tennessee 37996, USA
\end{affiliations}

\begin{abstract}
Quantum physics can be exploited to generate true random numbers, which play important roles in many applications, especially in cryptography. Genuine randomness from the measurement of a quantum system reveals the inherent nature of quantumness --- coherence, an important feature that differentiates quantum mechanics from classical physics. The generation of genuine randomness is generally considered impossible with only classical means. Based on the degree of trustworthiness on devices, quantum random number generators (QRNGs) can be grouped into three categories. The first category, practical QRNG, is built on fully trusted and calibrated devices and typically can generate randomness at a high speed by properly modeling the devices. The second category is self-testing QRNG, where verifiable randomness can be generated without trusting the actual implementation. The third category, semi-self-testing QRNG, is an intermediate category which provides a tradeoff between the trustworthiness on the device and the random number generation speed.
\end{abstract}

\maketitle

\section{Introduction}

Random numbers play essential roles in many fields, such as, cryptography\cite{Shannon1949OTP}, scientific simulations\cite{metropolis1949monte}, lotteries, and fundamental physics tests\cite{bell1964einstein}. These tasks rely on the unpredictability of random numbers, which generally cannot be guaranteed in classical processes. In computer science, random number generators (RNGs) are based on pseudo-random number generation algorithms\cite{knuth2014art}, which deterministically expand a random seed. Although the output sequences are usually perfectly balanced between 0s and 1s, a strong long-range correlation exists, which can undermine cryptographic security, cause unexpected errors in scientific simulations, or open loopholes in fundamental physics tests\cite{Kofler06,Hall10,Yuan2015CHSH}.

Many researchers have attempted to certify randomness solely based on the observed random sequences. In the 1950s, Kolmogorov developed the \emph{Kolmogorov complexity} concept to quantify the randomness in a certain string\cite{Kolmogorov1998387}. A RNG output sequence appears random if it has a high Kolmogorov complexity. Later, many other statistical tests\cite{marsaglia1996diehard,rukhin2001statistical,kim2004corrections} were developed to examine randomness in the RNG outputs. However, testing a RNG from its outputs can never prevent a malicious RNG from outputting a predetermined string that passes all of these statistical tests. Therefore, true randomness can only be obtained via processes involving inherent randomness.

\begin{figure}
\centering
\includegraphics[width=6cm]{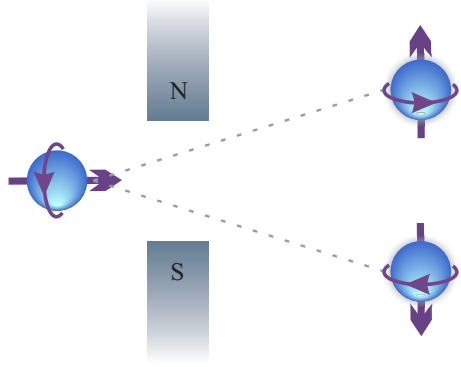}
\caption{Electron spin detection in the Stern-Gerlach experiment. Assume that the spin takes two directions along the vertical axis, denoted by $\ket{\uparrow}$ and $\ket{\downarrow}$. If the electron is initially in a superposition of the two spin directions, $\ket{\rightarrow} = (\ket{\uparrow} + \ket{\downarrow})/\sqrt{2}$, detecting the location of the electron would breaks the coherence and the outcome ($\uparrow$ or $\downarrow$) is intrinsically random.} \label{Fig:superposition}
\end{figure}

In quantum mechanics, a system can be prepared in a superposition of the (measurement) basis states, as shown in Fig.~\ref{Fig:superposition}. According to Born's rule, the measurement outcome of a quantum state can be intrinsically random, i.e. it can never be predicted better than blindly guessing. Therefore, the nature of inherent randomness in quantum measurements can be exploited for generating true random numbers.  Within a resource framework, coherence\cite{Baumgratz14} can be measured similarly to entanglement\cite{Bennett96}. By breaking the coherence or superposition of the measurement basis, it is shown that the obtained intrinsic randomness comes from the consumption of coherence. In turn, quantum coherence can be quantified from intrinsic randomness\cite{Yuan15}.

A practical QRNG can be developed using the simple process as shown in Fig.~\ref{Fig:superposition}. Based on the different implementations, there exists a variety of practical QRNGs. Generally, these QRNGs are featured for their high generation speed and a relatively low cost. In reality, quantum effects are always mixed with classical noises, which can be subtracted from the quantum randomness after properly modelling the underlying quantum process\cite{ma2013postprocessing}.

The randomness in the practical QRNGs usually suffices for real applications if the model fits the implementation adequately. However, such QRNGs can generate randomness with information-theoretical security only when the model assumptions are fulfilled. In the case that the devices are manipulated by adversaries, the output may not be genuinely random.
For example, when a QRNG is wholly supplied by a malicious manufacturer, who copies a very long random string to a large hard drive and only outputs the numbers from the hard drive in sequence, the manufacturer can always predict the output of the QRNG device.

On the other hand, a QRNG can be designed in a such way that its output randomness does not rely on any physical implementations. True randomness can be generated in a self-testing way even without perfectly characterizing the realisation instruments. The essence of a self-testing QRNG is based on device-independently witnessing quantum entanglement or nonlocality by observing a violation of the Bell inequality\cite{bell1964einstein}. Even if the output randomness is mixed with uncharacterised classical noise, we can still get a lower bound on the amount of genuine randomness based on the amount of nonlocality observed. The advantage of this type of QRNG is the self-testing property of the randomness. However, because the self-testing QRNG must demonstrate nonlocality, its generation speed is usually very low. As the Bell tests require random inputs, it is crucial to start with a short random seed. Therefore, such a randomness generation process is also called randomness expansion.

In general, a QRNG comprises a source of randomness and a readout system. In realistic implementations, some parts may be well characterised while others are not. This motivates the development of an intermediate type of QRNG, between practical and fully self-testing QRNGs, which is called semi-self-testing. Under several reasonable assumptions, randomness can be generated without fully characterising the devices. For instance, faithful randomness can be generated with a trusted readout system and an arbitrary untrusted randomness resource. A semi-self-testing QRNG provides a trade off between practical QRNGs (high performance and low cost) and self-testing QRNGs (high security of certified randomness).


In the last two decades, there have been tremendous development for all the three types of QRNG, trusted-device, self-testing, and semi-self-testing. In fact, there are commercial QRNG products available in the market. A brief summary of representative practical QRNG demonstrations that highlights the broad variety of optical QRNG is presented in Table \ref{Tab:SummaryPractical}. These QRNG schemes will be discussed further in Section \ref{Sec:Practical1} and \ref{Sec:Practical2}. A summary of self-testing and semi-self-testing QRNG demonstrations is presented in Table \ref{Tab:SummaryTheoretical}, which will be reviewed in details in Section \ref{Sec:Self} and \ref{Sec:Semi}.

\begin{table}
\centering
\caption{A brief summary of trusted-device QRNG demonstrations. Detailed description of these schemes can be found in Section \ref{Sec:Practical1} and \ref{Sec:Practical2}. Note that the quality/security of random numbers in different demonstrations may be different. Raw: reported raw generation rate, Refined: reported refined rate, Acquisition: data acquisition by dedicated hardware or commercial oscilloscope, SPD: single photon detector, BS: beam splitter, MCP-PCID: micro-channel-plate-based photon counting imaging detector, PNRD: photon-number-resolving detector, CMOS: complementary
metal-oxide-semiconductor, $-$: no related information found.} \label{Tab:SummaryPractical}
\begin{tabular}{ccccccc}
\hline
Year & Entropy source & Detection & Raw & Refined & Acquisition \\
  \hline
  2000 & Spatial mode\cite{jennewein2000fast} & SPD & 1 Mbps & $-$ & dedicated\\
  2000 & Spatial mode\cite{stefanov2000optical} & SPD & 100 Kbps & $-$ & dedicated \\
  2014& Spatial mode\cite{yan2014multi} & MCP-PCID & 8 Mbps& $-$ & dedicated \\
  2008& Temporal mode\cite{dynes2008high} & SPD & 4.01 Mbps & $-$ & dedicated \\
  2009& Temporal mode\cite{WJAK2009} & SPD & 55 Mbps & 40 Mbps & dedicated \\
  2011& Temporal mode\cite{WLB2011} & SPD & 180 Mbps & 152 Mbps & dedicated \\
  2014& Temporal mode\cite{nie2014practical} & SPD & 109 Mbps & 96 Mbps & dedicated \\
  2010& Photon number\cite{FWN2010} & PNRD  & 50 Mbps & $-$ & dedicated \\
  2011& Photon number\cite{Ren2011} & PNRD  & 2.4 Mbps & $-$ & dedicated \\
  2015& Photon number\cite{ATD2015} & PNRD  & $-$ & 143 Mbps & oscilloscope \\
  2010& Vacuum noise\cite{gabriel2010generator} & Homodyne  & 10 Mbps & 6.5 Mbps & dedicated \\
  2010& Vacuum noise\cite{Yong2010} & Homodyne  & $-$ & 12 Mbps & dedicated \\
  2011& Vacuum noise\cite{symul2011real} & Homodyne  & 3 Gbps & 2 Gbps & dedicated \\
  2010& ASE-intensity noise\cite{WSL2010} & Photo detector  & 12.5 Gbps  & $-$ & dedicated \\
  2011& ASE-intensity noise\cite{li2011scalable} & Photo detector  & 20 Gbps  & $-$ & $-$ \\
  2010& ASE-phase noise\cite{qi2010high} & Self-heterodyne  & 1 Gbps & 500 Mbps & oscilloscope \\
  2011& ASE-phase noise\cite{jofre2011true} & Self-heterodyne  & 1.2 Gbps & 1.11 Gbps & oscilloscope \\
  2012& ASE-phase noise\cite{xu2012ultrafast} & Self-heterodyne  & 8 Gbps & 6 Gbps & oscilloscope \\
  2014& ASE-phase noise\cite{YLD2014} & Self-heterodyne  & 80 Gbps & $-$ & oscilloscope \\
  2014& ASE-phase noise\cite{AAJ2014} & Self-heterodyne  & 82 Gbps & 43 Gbps & oscilloscope \\
  2015& ASE-phase noise\cite{nie201568} & Self-heterodyne  & 80 Gbps & 68 Gbps & oscilloscope \\
 \hline
\end{tabular}
\end{table}

\begin{table}
\centering
\caption{A summary of self-testing and semi-self-testing QRNG demonstrations. MDI: measurement device independent, SI: source independent, CV: continuous variable.} \label{Tab:SummaryTheoretical}
\begin{tabular}{ccccc}
\hline
Year & Type & Detection & Speed & Acquisition\\
  \hline
 2010 & Self-testing\cite{Pironio10} & ion-trap & very slow & dedicated \\
 2013 & Self-testing\cite{giustina2013bell} & SPD & 0.4 {bps}& dedicated \\
 2015& SI\cite{CAO2016SOURCE}&SPD&5 Kbps & dedicated \\
 2015& CV-SI\cite{2015arXiv150907390M} & Homodyne & 1 Gbps& oscilloscope \\
 2015& Self-testing with fixed dimension\cite{Lunghi15}&SPD& 23 bps& dedicated \\
 \hline
\end{tabular}
\end{table}

\section{Trusted-device QRNG I: single-photon detector} \label{Sec:Practical1}
True randomness can be generated from any quantum process that breaks coherent superposition of states. Due to the availability of high quality optical components and the potential of chip-size integration, most of today's practical QRNGs are implemented in photonic systems. In this survey, we focus on various implementations of optical QRNGs.

A typical QRNG includes an entropy source for generating well-defined quantum states and a corresponding detection system. The inherent quantum randomness in the output is generally mixed with classical noises.  Ideally, the extractable quantum randomness should be well quantified and be the dominant source of the randomness. By applying randomness extraction, genuine randomness can be extracted from the mixture of quantum and classical noise. The extraction procedure is detailed in Methods.

\subsection{Qubit state}
\begin{figure}
\centering
\includegraphics[width=12 cm]{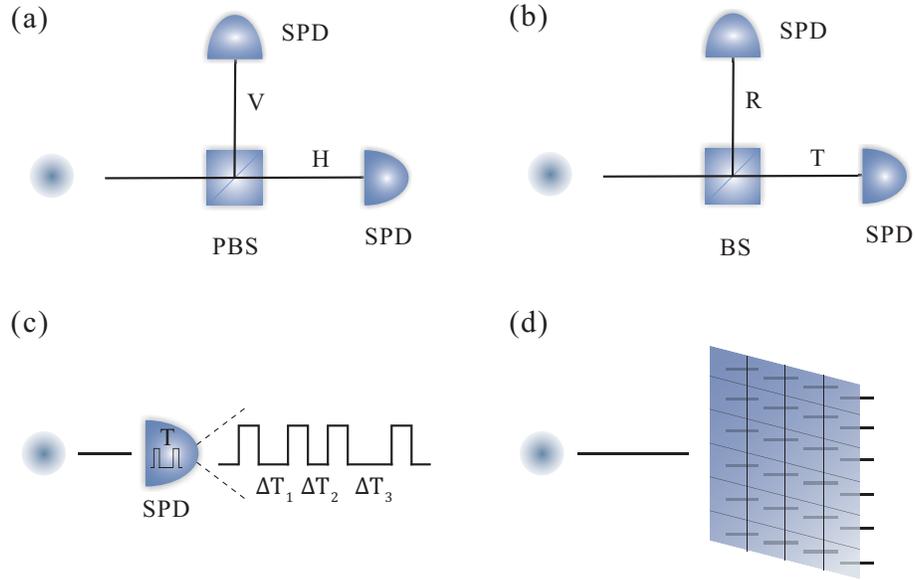}
\caption{Practical QRNGs based on single photon measurement. (a) A
photon is originally prepared in a superposition of horizontal (H) and vertical (V) polarizations, described by $(\ket{H}+\ket{V})/\sqrt{2}$. A polarising beam splitter (PBS) transmits the horizontal and reflects the vertical polarization. For random bit generation, the photon is measured by two single photon detectors (SPDs). (b) After passing through a symmetric beam splitter (BS), a photon exists in a superposition of transmitted (T) and reflected (R) paths, $(\ket{R}+\ket{T})/\sqrt{2}$. A random bit can be generated by measuring the path information of the photon. (c) QRNG based on measurement of photon arrival time. Random bits can be generated, for example, by measuring the time interval, $\Delta t$, between two detection events. (d) QRNG based on measurements of photon spatial mode. The generated random number depends on spatial position of the detected photon, which can be read out by an SPD array.} \label{Fig:SinglePhoton}
\end{figure}

Random bits can be generated naturally by measuring a qubit\footnote{A qubit is a two-level quantum-mechanical system, which, similar to a bit in classical information theory, is the fundamental unit of quantum information.} $\ket{+}=(\ket{0}+\ket{1})/\sqrt2$ in the $Z$ basis, where $\ket{0}$ and $\ket{1}$ are the eigenstates of the measurement $Z$. For example, Fig.~\ref{Fig:SinglePhoton}~(a) shows a polarization based QRNG, where $\ket{0}$ and $\ket{1}$ denote horizontal and vertical polarization, respectively, and $\ket{+}$ denotes $+45^o$ polarization. Fig.~\ref{Fig:SinglePhoton}~(b) presents a path based QRNG, where $\ket{0}$ and $\ket{1}$ denote the photon traveling via path $R$ and $T$, respectively.

The most appealing property of this type of QRNGs lies on their simplicity in theory that the generated randomness has a clear quantum origin. This scheme was widely adopted in the early development of QRNGs\cite{rarity1994quantum,stefanov2000optical,jennewein2000fast}. Since at most one random bit can be generated from each detected photon, the random number generation rate is limited by the detector's performance, such as dead time and efficiency. For example, the dead time of a typical silicon SPD based on an avalanche diode is tens of ns\cite{Eisaman2011PD}. Therefore, the random number generation rate is limited to tens of Mbps, which is too low for certain applications such as high-speed quantum key distribution (QKD), which can be operated at GHz clock rates\cite{takesue2007quantum,PhysRevX.2.041010}. Various schemes have been developed to improve the performance of QRNG based on SPD.

\subsection{Temporal mode}
One way to increase the random number generation rate is to perform measurement on a high-dimensional quantum space, such as measuring the temporal or spatial mode of a photon. Temporal QRNGs measure the arrival time of a photon, as shown in Fig.~\ref{Fig:SinglePhoton}~(c). In this example, the output of a continuous-wave laser is detected by a time-resolving SPD. The laser intensity can be carefully controlled such that within a chosen time period $T$, there is roughly one detection event. The detection time is randomly distributed within the time period $T$ and digitized with a time resolution of $\delta_t$. The time of each detection event is recorded as raw data. Thus for each detection, the QRNG generates about $\log_2(T/\delta_t)$ bits of raw random numbers. Essentially, $\delta_t$ is limited by the time jitter of the detector (typically in the order of 100 ps), which is normally much smaller than the detector deadtime (typically in the order of 100 ns)\cite{Eisaman2011PD}.

One important advantage of temporal QRNGs is that more than one bit of random number can be extracted from a single-photon detection, thus improving the random number generation rate. The time period $T$ is normally set to be comparable to the detector deadtime. Comparing to the qubit QRNG, the temporal-mode QRNG alleviates the impact of detection deadtime. For example, if the time resolution and the dead time of an SPD are 100 ps and 100 ns respectively, the generation rate of temporal QRNG is around $\log_2(1000)\times$10 Mbps, which is higher than that of the qubit scheme (limited to 10 Mbps). The temporal QRNGs have been well studied recently\cite{ma2005random,dynes2008high,WJAK2009,WLB2011,nie2014practical}.

\subsection{Spatial mode}
Similar to the case of temporal QRNG, multiple random bits can be generated by measuring the spatial mode of a photon with a space-resolving detection system. One illustrative example is to send a photon through a $1\times N$ beam splitter and to detect the position of the output photon. Spatial QRNG has been experimentally demonstrated by using a multi-pixel single-photon detector array\cite{yan2014multi}, as shown in Fig.~\ref{Fig:SinglePhoton}~(d).  The distribution of the random numbers depends on both the spatial distribution of light intensity and the efficiency uniformity of the SPD arrays.

The spatial QRNG offers similar properties as the temporal QRNG, but requires multiple detectors. Also, correlation may be introduced between the random bits because of cross talk between different pixels in the closely-packed detector array.

\subsection{Multiple photon number states}
Randomness can be generated not only from measuring a single photon, but also from quantum states containing multiple photons. For instance, a coherent state
\begin{equation}\label{}
\ket{\alpha} = e^{-\frac{|\alpha|^2}{2}}\sum_{n=0}^\infty \frac{\alpha^n}{\sqrt{n!}}\ket{n},
\end{equation}
is a superposition of different photon-number (Fock) states $\{\ket{n}\}$, where $n$ is the photon number and $|\alpha|^2$ is the mean photon number of the coherent state. Thus, by measuring the photon number of a coherent laser pulse with a photon-number resolving SPD, we can obtain random numbers that follow a Poisson distribution.  QRNGs based on measuring photon number have been successfully demonstrated in experiments\cite{FWN2010,Ren2011,ATD2015}. Interestingly, random numbers can be generated by resolving photon number distribution of a light-emitting diode (LED) with a consumer-grade camera inside a mobile phone, as shown in a recent study\cite{Mobile2014}.

Note that, the above scheme is sensitive to both the photon number distribution of the source and the detection efficiency of the detector. In the case of a coherent state source, if the loss can be modeled as a beam splitter, the low detection efficiency of the detector can be easily compensated by using a relatively strong laser pulse.

\section{Trusted-device QRNG II: macroscopic photodetector} \label{Sec:Practical2}
The performance of an optical QRNG largely depends on the employed detection device. Beside SPD, high-performance macroscopic photodetectors have also been applied in various QRNG schemes. This is similar to the case of QKD, where protocols based on optical homodyne detection\cite{grosshans2003quantum} have been developed, with the hope to achieve a higher key rate over a low-loss channel. In the following discussion, we review two examples of QRNG implemented with macroscopic photodetector.

\subsection{Vacuum noise}
\begin{figure}[!hbt]
\centering
\includegraphics[width=11 cm]{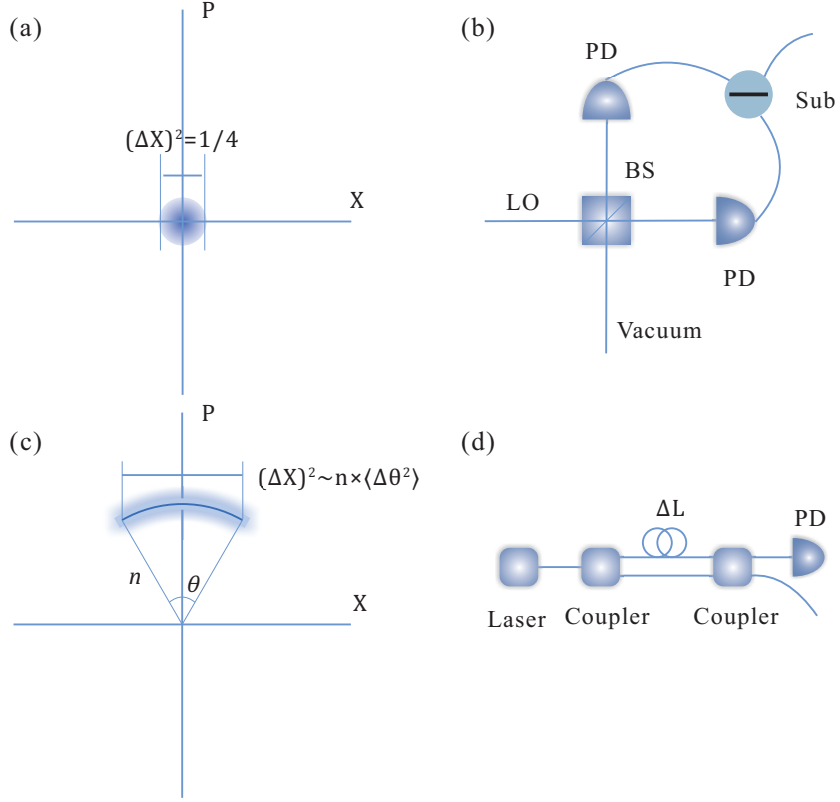}
\caption{QRNGs using macroscopic photodetector. (a) Phase-space representation of the vacuum state. The variance of the $X$-quadrature is 1/4. (b) QRNG based on vacuum noise measurements. The system comprises a strong local oscillator (LO), a symmetric beam splitter (BS), a pair of photon detector (PD), and an electrical subtracter (Sub).  (c) Phase-space representation of a partially phase-randomised coherent state. The variance of the $X$-quadrature is in the order of $n\times\langle\Delta\theta^2\rangle$, where $n$ is the average photon number and $\langle\Delta\theta^2\rangle$ is the phase noise variance. (d) QRNGs based on measurements of laser phase noise. The first coupler splits the original laser beam into two beams, which propagate through two optical fibres of different lengths, thereafter interfering at the second coupler. The output signal is recorded by a photon detector. The extra length $\Delta L$ in one fibre introduces a time delay $T_d$ between the two paths, which in turn determines the variance of the output signal.}
\label{Fig:Vac}
\end{figure}

In quantum optics, the amplitude and phase quadratures of the vacuum state are represented by a pair of non-commuting operators ($X$ and $P$ with $[X,P]=i/2$), which cannot be determined simultaneously with an arbitrarily high precision\cite{Braunstein05}, i.e. $\langle(\Delta X)^2\rangle\times\langle(\Delta P)^2\rangle\ge1/16$, with $\Delta O$ defined by $O - \langle O\rangle$ and $\langle O\rangle$ denoting the average of $O$. This can be easily visualised in the phase space, where the vacuum state is represented by a two-dimensional Gaussian distribution centered at the origin with an uncertainty of $1/4$ (the shot-noise variance) along any directions, as shown in Fig.~\ref{Fig:Vac}~(a). In principle, Gaussian distributed random numbers can be generated by measuring any field quadrature repeatedly. This scheme has been implemented by sending a strong laser pulse
through a symmetric beam splitter and detecting the differential signal of the two output beams with a balanced receiver\cite{gabriel2010generator,Yong2010,symul2011real}.

Given that the local oscillator (LO) is a single-mode coherent state and the detector is shot-noise limited, the random numbers generated in this scheme follow a Gaussian distribution, which is on demand in certain applications, such as Gaussian-Modulated Coherent States (GMCS) QKD\cite{grosshans2003quantum}. There are several distinct advantages of this approach. First, the resource of quantum randomness, the vacuum state, can be easily prepared with a high fidelity. Second, the performance of the QRNG is insensitive to detector loss, which can be simply compensated by increasing the LO power. Third, the field quadrature of vacuum is a continuous variable, suggesting that more than one random bit can be generated from one measurement. For example, 3.25 bits of random numbers are generated from each measurement\cite{gabriel2010generator}.

In practice, an optical homodyne detector itself contributes additional technical noise, which may be observed or even controlled by a potential adversary. A randomness extractor is commonly required to generate secure random numbers. To extract quantum randomness effectively, the detector should be operated in the shot-noise limited region, in which the overall observed noise is dominated by vacuum noise. We remark that building a broadband shot-noise limited homodyne detector operating above a few hundred MHz is technically challenging\cite{okubo2008pulse,chi2011balanced,kumar2012versatile}. This may in turn limit the ultimate operating speed of this type of QRNG.

\subsection{Amplified spontaneous emission}
To overcome the bandwidth limitation of shot-noise limited homodyne detection, researchers have developed QRNGs based on measuring phase\cite{qi2010high,jofre2011true,xu2012ultrafast,AAJ2014,YLD2014,nie201568} or intensity noise\cite{WSL2010,li2011scalable} of amplified spontaneous emission(ASE), which is quantum mechanical by nature\cite{henry1982theory,ma2013postprocessing,zhou2015randomness}.

In the phase-noise based QRNG scheme, random numbers are generated by measuring a field quadrature of \emph{phase-randomized} weak coherent states (signal states). Figure \ref{Fig:Vac}~(c) shows the phase-space representation of a signal state with an average photon number of $n$ and a phase variance of $\langle(\Delta\theta)^2\rangle$. If the average phase of the signal state is around $\pi/2$, the uncertainty of the $X$-quadrature is of the order of  $n\langle(\Delta\theta)^2\rangle$. When $n$ is large, this uncertainty can be significantly larger than the vacuum noise. Therefore, phase noise based QRNG is more robust against detector noise. In fact, this scheme can be implemented with commercial photo-detectors operated above GHz rates.

QRNG based on laser phase noise was first developed using a cw laser source and a delayed self-heterodyning detection system\cite{qi2010high}, as shown in Fig.~\ref{Fig:Vac}~(d). Random numbers are generated by measuring the phase difference of a single-mode laser at times $t$ and $t+T_d$. Intuitively, if the time delay $T_d$ is much larger than the coherence time of the laser, the two laser beams interfering at the second beam splitter can be treated as generated by independent laser sources. In this case, the phase difference is a random variable uniformly distributed in $[-\pi, \pi)$, regardless of the classical phase noise introduced by the unbalanced interferometer itself. This suggests that a robust QRNG can be implemented without phase-stabilizing the interferometer. On the other hand, by phase-stabilizing the interferometer, the time delay $T_d$ can be made much shorter than the coherent time of the laser\cite{qi2010high}, enabling a much higher sampling rate. This phase stabilization scheme has been adopted in a $\ge6$ Gbps QRNG\cite{xu2012ultrafast} and a 68 Gbps QRNG demonstration\cite{nie201568}.

Phase noise based QRNG has also been implemented using pulsed laser source, where the phase difference between adjacent pulses is automatically randomized\cite{jofre2011true,AAJ2014,YLD2014}. A speed of 80 Gbps (raw rate as shown in Table1) has been demonstrated\cite{YLD2014}. It also played a crucial role in a recent loophole-free Bell experiment\cite{AAM2015}. Here, we want to emphasize that strictly speaking, none of these generation speeds are real-time, due to the speed limitation of the randomness extraction\cite{ma2013postprocessing}. Although such limitation is rather technical, in practice, it is important to develop extraction schemes and hardware that can match the fast random bit generation speed in the future.

\section{Self-testing QRNG} \label{Sec:Self}
Realistic devices inevitably introduce classical noise that affects the output randomness, thus causing the generated random numbers depending on certain classical variables, which might open up security issues. To remove this bias, one must properly model the devices and quantify their contributions. In the QRNG schemes described in Section \ref{Sec:Practical1} and Section \ref{Sec:Practical2}, the output randomness relies on the device models\cite{ma2013postprocessing,zhou2015randomness}. When the implementation devices deviate from the theoretical models, the randomness can be compromised. In this section, we discuss \emph{self-testing} QRNGs, whose output randomness is certified independent of device implementations.

\subsection{Self-testing randomness expansion}
\begin{figure*}[hbt]
\centering
\resizebox{4cm}{!}{\includegraphics{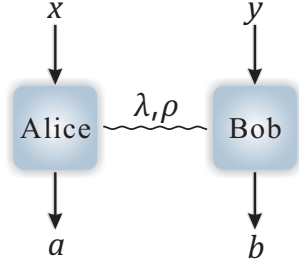}}
\caption{Illustration of a bipartite Bell test. Alice and Bob are two spacelikely separated parties, that output $a$ and $b$ from random inputs $x$ and $y$, respectively. A Bell inequality is defined as a linear combination of the probabilities $p(a,b|x,y)$. For instance, the Clauser-Horne-Shimony-Holt (CHSH) inequality\cite{CHSH} is defined by $S = \sum_{a,b,x,y} (-1)^{a + b + xy}p(a,b|x,y) \leq S_C = 2$, where all of the inputs and outputs are bit values, and $S_C$ is the classical bound for all local hidden-variable models. With quantum settings, that is, performing measurements $M_x^a\otimes M_y^b$ on quantum state $\rho_{AB}$, $p(a,b|x,y) = \mathrm{Tr}[\rho_{AB}M_x^a\otimes M_y^b]$, the CHSH inequality can be violated up to $S_Q = 2\sqrt{2}$. Quantum features (such as intrinsic randomness) manifest as violations of the CHSH inequality.} \label{Fig:Bell}
\end{figure*}

In QKD, secure keys can be generated even when the experimental devices are not fully trusted or characterised\cite{Mayers98,acin06}. Such self-testing processing of quantum information  also occur in randomness generation (expansion). The output randomness can be certified by observing violations of the Bell inequalities\cite{bell1964einstein}, see Fig.~\ref{Fig:Bell}. Under the no-signalling condition\cite{prbox} in the Bell tests, it is impossible to violate Bell inequalities if the output is not random, or, predetermined by local hidden variables.

Since Colbeck\cite{Colbeck09, Colbeck11} suggested that randomness can be expanded by untrusted devices, several protocols based on different assumptions have been proposed. For instance, in a non-malicious device scenario, we can consider that the devices are honestly designed but get easily corrupt by unexpected classical noises. In this case, instead of a powerful adversary that may entangle with the experiment devices, we can consider a classical adversary who possesses only classical knowledge of the quantum system and analyzes the average randomness output conditioned by the classical information. Based on the Clauser-Horne-Shimony-Holt (CHSH) inequality\cite{CHSH}, Fehr et al.\cite{Fehr13} and Pironio et al.\cite{Pironio13} proposed self-testing randomness expansion protocols against classical adversaries. The protocols quadratically expands the input seed, implying that the length of the input seed is $O(\sqrt{n}\log_2\sqrt{n})$, where $n$ denotes the experimental iteration number.

A more sophisticated exponential randomness expansion protocol based on the CHSH inequality was proposed by Vidick and Vazirani\cite{Vazirani12}, in which the lengths of the input seed is $O(\log_2n)$. In the same work, they also presented an exponential expansion protocol against quantum adversaries, where quantum memories in the devices may entangle with the adversary. The Vidick-Vazirani protocol against quantum adversaries places strict requirements on the experimental realisation. Miller and Shi\cite{Miller14} partially solved this problem by introducing a more robust protocol. Combined with the work by Chung, Shi, and Wu\cite{Chung14}, they also presented an unbounded randomness expansion scheme. By adopting a more general security proof, Miller and Shi\cite{Miller15} recently showed that genuinely randomness can be obtained as long as the CHSH inequality is violated. Their protocol greatly improves the noise tolerance, indicating that an experimental realisation of a fully self-testing randomness expansion protocol is feasible.

The self-testing randomness expansion protocol relies on a faithful realisation of Bell test excluding the experimental loopholes, such as locality and efficiency loopholes. The randomness expansion protocol against classical adversaries is firstly experimentally demonstrated by Pironio et al.\cite{Pironio10} in an ion-trap system, which closes the efficiency loophole but not the locality loophole. To experimentally close the locality loophole, a photonic system is more preferable when quantum memories are unavailable. As the CHSH inequality is minimally violated in an optically realised system\cite{giustina2013bell,Christensen13}, the randomness output is also very small (with min-entropy of $H_{\mathrm{min}} = 7.2\times10^{-5}$ in each run), and the randomness generation rate is $0.4$ {bits/s}. To maximise the output randomness, the implementation settings are designed to maximally violate the CHSH inequality. Due to experimental imperfections, the chosen Bell inequality might be sub-optimal for the observed data. In this case, the output randomness can be optimised over all possible Bell inequalities\cite{Silleras14,Bancal14b}.

Although nonlocality or entanglement certifies the randomness, the three quantities, nonlocality, entanglement, and randomness are not equivalent\cite{acin12}. Maximum randomness generation does not require maximum nonlocal correlation or a maximum entangled state. In the protocols based on the CHSH inequality, maximal violation (nonlocality and entanglement) generates 1.23 bits of randomness. It is shown that 2 bits of randomness can be certified with little involvement of nonlocality and entanglement\cite{acin12}. Furthermore, as discussed in a more generic scenario involving nonlocality and randomness, it is shown that maximally nonlocal theories cannot be maximally random\cite{Torre15}.

\subsection{Randomness amplification}
In self-testing QRNG protocols based on the assumption of perfectly random inputs, the output randomness is guaranteed by the violations of Bell tests. Conversely, when all the inputs are predetermined, any Bell inequality can be violated to an arbitrary feasible value without invoking a quantum resource. Under these conditions, all self-testing QRNG protocols cease to work any more. Nevertheless, randomness generation in the presence of partial randomness is still an interesting problem. Here, an adversary can use the additional knowledge of the inputs to fake violations of Bell inequalities. The task of generating arbitrarily free randomness from partially free randomness is also called randomness amplification, which is impossible to achieve in classical processes.

The first randomness amplification protocol was proposed by Colbeck and Renner\cite{Colbeck12}. Using a two-party chained Bell inequality\cite{Pearle70,Braunstein90}, they showed that any Santha-Vazirani weak sources\cite{santha1986generating} (defined in Methods), with $\epsilon<0.058$, can be amplified into arbitrarily free random bits in a self-testing way by requiring only no-signaling. A basic question of randomness amplification is whether free random bits can be obtained from arbitrary weak randomness. This question was answered by Gallego et al.\cite{Gallego13}, who demonstrated that perfectly random bits can be generated using a five-party Mermin inequality\cite{Mermin90} with arbitrarily imperfect random bits under the no-signaling assumption.

Randomness amplification is related to the freewill assumption\cite{Kofler06, Hall10,Barrett11,Koh12,Pope13,putz14,Yuan2015CHSH} in Bell tests. In experiments, the freewill assumption requires the inputs to be random enough such that violations of Bell inequalities are induced from quantum effects rather than predetermined classical processes. This is extremely meaningful in fundamental Bell tests, which aim to rule out local realism. Such fundamental tests are the foundations of self-testing tasks, such as device-independent QKD and self-testing QRNG. Interestingly, self-testing tasks require a faithful violation of a Bell inequality, in which intrinsic random numbers are needed. However, to generate faithful random numbers, we in turn need to witness nonlocality which requires additional true randomness. Therefore, the realisations of genuine loophole-free Bell tests and, hence, fully
self-testing tasks are impossible. Self-testing protocols with securities independent of the untrusted part can be designed only by placing reasonable assumptions on the
trusted part.

\section{Semi-self-testing QRNGs} \label{Sec:Semi}
Traditional QRNGs based on specific models pose security risks in fast random number generation. On the other hand, the randomness generated by self-testing QRNGs is information-theoretically secure even without characterising the devices, but the processes are impractically slow. As a compromise, intermediate QRNGs might offer a good tradeoff between trusted and self-testing schemes --- realising both reasonably fast and secure random number generation.

\begin{figure}[hbt]
\centering
\resizebox{5cm}{!}{\includegraphics{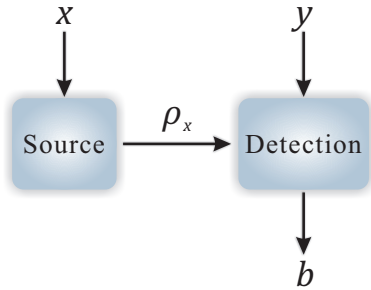}}
\caption{A semi-self-testing QRNG. Conditional on the input setting $x$, the source emits a quantum state $\rho_x$. Conditional on the input $y$, the detection device measures $\rho_x$ and outputs $b$. } \label{Fig:Semi}
\end{figure}

As shown in Fig.~\ref{Fig:Semi}, a typical QRNG comprises two main modules, a source that emits quantum states and a measurement device that detects the states and outputs random bits.  In trusted-device QRNGs, both source and measurement devices\cite{ma2013postprocessing,zhou2015randomness} must be modeled properly; while the output randomness in the fully self-testing QRNGs does not depend on the implementation devices.

In practice, there exist scenarios that the source (respectively, measurement device) is well characterised, while the measurement device (respectively, source) not. Here, we review the semi-self-testing QRNGs, where parts of the devices are trusted.

\subsection{Source-independent QRNG}
In source-independent QRNG, the randomness source is assumed to be untrusted, while the measurement devices are trusted. The essential idea for this type of scheme is to use the measurement to monitor the source in real time. In this case, normally one needs to randomly switch among different (typically, complement) measurement settings, so that the source (assumed to be under control of an adversary) cannot predict the measurement ahead. Thus, a short seed is required for the measurement choices.

In the illustration of semi-self-testing QRNG, Fig.~\ref{Fig:Semi}, the source-independent scheme is represented by a unique $x$ (corresponding to a state $\rho_x$) and multiple choices of the measurement settings $y$. In Section \ref{Sec:Practical1}, we present that randomness can be obtained by measuring $\ket{+}$ in the $Z$ basis. However, in a source-independent scenario, we cannot assume that the source emits the state $\ket{+}$. In fact, we cannot even assume the dimension of the state $\rho_x$. This is the major challenge facing for this type of scheme.

In order to faithfully quantify the randomness in the $Z$ basis measurement, first a squashing model is applied so that the to-be-measured state is equivalent to a qubit\cite{BML_Squash_08}. Note that this squashing model puts a strong restriction on measurement devices. Then, the measurement device should occasionally project the input state onto the $X$ basis states, \ket{+} and \ket{-}, and check whether the input is $\ket{+}$\cite{CAO2016SOURCE}. The technique used in the protocol shares strong similarity with the one used in QKD\cite{Shor2000Simple}. The $X$ basis measurement can be understood as the \emph{phase error estimation}, from which we can estimate the amount of classical noise. Similar to privacy amplification, randomness extraction is performed to subtract the classical noise and output true random values.


The source-independent QRNG is advantageous when the source is complicated, such as in the aforementioned QRNG schemes based on measuring single photon sources\cite{stefanov2000optical,jennewein2000fast,rarity1994quantum}, LED lights\cite{Mobile2014}, and phase fluctuation of lasers\cite{xu2012ultrafast}. In these cases, the sources are quantified by complicated or hypothetical physical models. Without a well-characterized source, randomness can still be generated. The disadvantage of this kind of QRNGs compared to fully self-testing QRNGs is that they need a good characterization of the measurement devices. For example, the upper and the lower bounds on the detector efficiencies need to be known to avoid potential attacks induced from detector efficiency mismatch. Also the intensity of light inputs into the measurement device needs to be carefully controlled to avoid attacks on the detectors.

Recently,  a continuous-variable version of the source-independent QRNG is experimentally demonstrated\cite{2015arXiv150907390M}
and achieves a randomness generation rate over 1 Gbps. Moreover, with state-of-the-art devices, it can potentially reach the speed in the order of tens of Gbps, which is similar to the trusted-device QRNGs. Hence, semi-self-testing QRNG is approaching practical regime.

\subsection{Measurement-device-independent QRNGs}
Alternatively, we can consider the scenario that the input source is well characterised while the measurement device is untrusted. In Fig.~\ref{Fig:Semi}, different inputs $\rho_x$ (hence multiple $x$) are needed to calibrate the measurement device with a unique setting $y$. Similar to the source-independent scenario, the randomness is originated by measuring the input state $\ket{+}$ in the $Z$ basis. The difference is that here the trusted source sends occasionally auxiliary quantum states $\rho_x$, such as $\ket{0}$, to check whether the measurement is in the $Z$ basis\cite{MDIQRNG15}.  The analysis combines measurement tomography with randomness quantification of positive-operator valued measure, and does not assume to know the dimension of the measurement device, i.e., the auxiliary ancilla may have an arbitrary dimension.

The advantage of such QRNGs is that they remove all detector side channels, but the disadvantage is that they may be subject to imperfections in the modeling of the source.  This kind of QRNG is complementary to the source-independent QRNG, and one should choose the proper QRNG protocol based on the experimental devices.

We now turn to two variations of measurement-device-independent QRNGs. First, the measurement tomography step may be replaced by a certain witness, which could simplify the scheme at the expense of a slightly worse performance. Second, similar to the source-independent case, a continuous-variable version of measurement-device-independent QRNG might significantly increase the bit rate. The challenge lies on continuous-variable entanglement witness and measurement tomography.

\subsection{Other semi-self-testing QRNGs}
Apart from the above two types of QRNGs, there are also some other QRNGs that achieve self-testing except under some mild assumptions. For example, the source and measurement devices can be assumed to occupy independent two-dimensional quantum subspaces\cite{Lunghi15}. In this scenario, the QRNG should use both different input states and different measurement settings. The randomness can be estimated by adopting a \emph{dimension witness}\cite{PRL.112.140407}. A positive value of this dimension witness could certify randomness in this scenario, similar to the fact that a violation of the Bell inequality could certify randomness of self-testing QRNG in Section \ref{Sec:Self}.

\section{Outlook}
The needs of ``perfect'' random numbers in quantum communication and fundamental physics experiments have stimulated the development of various QRNG schemes, from highly efficient systems based on trusted devices, to the more theoretically interesting self-testing protocols. On the practical side, the ultimate goal is to achieve fast random number generation at low cost, while maintaining high-level of randomness. With the recent development on waveguide fabrication technique\cite{barak2003true}, we expect that chip-size, high-performance QRNGs could be available in the near future. In order to guarantee the output randomness, the underlying physical models for these QRNGs need to be accurate and both the quantum noise and classical noise should be well quantified. Meanwhile, by developing a semi-self-testing protocol, a QRNG becomes more robust against classical noises and device imperfections. In the future, it is interesting to investigate the potential technologies required to make the self-testing QRNG practical. With the new development on single-photon detection, the readout part of the self-testing QRNG can be ready for practical application in the near future. The entanglement source, on the other hand, is still away from the practical regime (Gbps).

On the theoretical side, the study of self-testing QRNG has not only provided means of generating robust randomness, but also greatly enriched our understanding on the fundamental questions in physics. In fact, even in the most recent loophole-free Bell experiment\cite{hensen2015experimental, PhysRevLett.115.250402, PhysRevLett.115.250401,ballance2015hybrid} where high-speed QRNG has played a crucial role, it is still arguable whether it is appropriate to use randomness generated based on quantum theory to test quantum physics itself. Other random resources have also been proposed for loophole-free Bell's inequality tests, such as independent comic photons\cite{Gallicchio14}. It is an open question whether we can go beyond QRNG and generate randomness from a more general theory.




\section*{Methods}
\subsection{Min-entropy source}
Given the underlying probability distribution, the randomness of a random sequence $X$ on $\{0,1\}^n$ can be quantified by its \emph{min-entropy}
\begin{equation}\label{}
H_{\mathrm{min}} = -\log\left(\max_{v\in\{0,1\}^n}\mathrm{Prob}[X=v]\right).
\end{equation}

\subsection{Santha-Vazirani weak sources\cite{santha1986generating}}
We assume that random bit numbers are produced in the time sequence $x_1, x_2, ..., x_j, ...$. Then, for $0<\epsilon\le 1/2$, the source is called $\epsilon$-free if
\begin{equation}\label{}
  \epsilon \le P(x_j|x_1, x_2, \dots, x_{j-1}, e) \le 1-\epsilon,
\end{equation}
for all values of $j$. Here $e$ represents all classical variables generated outside the future light-cone of the Santha-Vazirani weak sources.

\subsection{Randomness extractor}
A RNG typically consists of two components, an entropy source and a randomness extractor\cite{barak2003true}. In a QRNG, the entropy source could be a physical device whose output is fundamentally unpredictable, while the randomness extractor could be an algorithm that generates nearly perfect random numbers from the output of the above preceding entropy source, which can be imperfectly random. The two components of QRNG are connected by quantifying the randomness with min-entropy. The min-entropy of the entropy source is first estimated and then fed into the randomness extractor as an input parameter.

The imperfect randomness of the entropy source can already be seen in the SPD based schemes, such as the photon number detection scheme. By denoting $N$ as the discrimination upper bound of a photon number resolving detector, at most $log_2(N)$ raw random bits can be generated per detection event. However, as the photon numbers of a coherent state source follows a Poisson distribution, the raw random bits follow a non-uniform distribution; consequently, we cannot obtain $log_2(N)$ bits of random numbers. To extract perfectly random numbers, we require a postprocessing procedure (i.e. randomness extractor).

In the coherent detection based QRNG, the quantum randomness is inevitably mixed with classical noises introduced by the detector and other system imperfections. Moreover, any measurement system has a finite bandwidth, implying unavoidable correlations between adjacent samples. Once quantified, these unwanted side-effects can be eliminated through an appropriate randomness extractor\cite{ma2013postprocessing}.

The composable extractor was first introduced in classical cryptography\cite{canetti2001universally,canetti2002universally}, and was later extended to quantum cryptography\cite{ben2005universal,renner2005universally}. To generate information-theoretically provable random numbers, two typical extractor, the Trevisan's extractor or the
Toeplitz-hashing extractor, are generally employed in practice.

Trevisan's extractor\cite{trevisan2001extractors,raz1999extracting} has been proven secure against quantum adversaries\cite{de2012trevisan}. Moreover, it is a strong extractor (its seed can be reused) and its seed length is polylogarithmic function of the input. Tevisan's extractor comprises two main parts, a one-bit extractor and a combinatorial design. The Toeplitz-hashing extractor was well developed in the privacy amplification procedure of the QKD system\cite{uchida2008fast}. This kind of extractor is also a strong extractor\cite{wegman1981new}. By applying the fast Fourier transformation technique, the runtime of the Toeplitz-hashing extractor can be improved to $O(n\log n)$.

On account of their strong extractor property, both of these extractors generate random numbers even when the random seed is longer than the output length of each run. Both extractors have been implemented\cite{ma2013postprocessing} and the speed of both extractors have been increased in follow-up studies\cite{DBLP:journals/corr/abs-1212-0520,ma2011explicit}, but remain far below the operating speed of the QRNG based on laser-phase fluctuation (68 Gbps\cite{nie201568}). Therefore, the speed of the extractor is the main limitation of a practical QRNG.

\section*{Acknowledgements}
The authors thank R.~Colbeck, H.-K.~Lo, Y.~Shi, and F.~Xu for enlightening discussions. This work was supported by the National Basic Research Program of China Grants No.~2011CBA00300 and No.~2011CBA00301, the 1000 Youth Fellowship program in China, and the Laboratory Directed Research and Development (LDRD) Program of Oak Ridge National Laboratory (managed by UT-Battelle LLC for the U.S.~Department of Energy).

\section*{Author contributions}
All authors contributed extensively to the work presented in this paper. X.~Y.~focused on the self-testing QRNG part. Z.~C.~focused on the semi-self-testing QRNG part. B.~Q.~and Z.~Z.~focused on the practical QRNG part. X.~M.~supervised the project.

\section*{Competing financial interests}
The authors declare no competing financial interests.

\bibliographystyle{naturemag}
\bibliography{BibQRNGNPJ}

\end{document}